\documentstyle[epsfig]{aipproc}

\begin{document}
\title{Evolving Stellar Background Radiation and Gamma-Ray Optical Depth}

\author{Tanja M. Kneiske$^*$, Karl Mannheim$^*$ and Dieter Hartmann$^{\dagger}$}
\address{$^*$Universit\"ats-Sternwarte G\"ottingen \\
Geismarlandstrasse 11, DE-37083 G\"ottingen, Germany\\
$^{\dagger}$Clemson University, Clemson, SC 29634-0978, USA}

\maketitle

\begin{abstract}
We present a semi-empirical model for the evolving far-infrared to 
ultraviolet diffuse background produced by stars in galaxies. 
The model is designed to reproduce the results of deep galaxy surveys, 
and therefore may be considered as a cosmology-independent lower limit to the 
extragalactic background light. Using this model and recent
HEGRA data,
we infer the intrinsic spectrum at multi-TeV gamma-ray energies for Mkn~501 and
find that it is consistent with a power law of spectral index
$2.49\pm0.04$.  In turn, this finding renders it rather unlikely that the
present-day infrared background has an intensity as high as claimed by 
Finkbeiner et al.~\cite{lit:finkbeiner}.
Future 10~GeV to TeV observations
could be used to either constrain the ultraviolet-to-infrared background model
at high redshifts or cosmological parameters.
\end{abstract}

\section*{Introduction}
High-energy gamma-rays originating from sources at cosmological distances 
can be absorbed by pair production in collisions with low-energy photons. Most
of the low-energy photons are produced by stars in ordinary galaxies and
contribute to the
extragalactic background light (EBL).  The role of other contributers,
such as Active Galactic Nuclei, is under dispute, and can safely be
assumed to be unimportant in terms of the global energy budget.  
To model the stellar component of the EBL, we use an
approach similar to that used in Refs.
\cite{lit:salsteck,lit:malsteck,lit:kneiske1}, however
with the star formation rate (SFR)
inferred from deep multi-color galaxy surveys. Other approaches
aim at finding an {\em ab initio} description of the 
SFR from the theory of structure formation \cite{lit:somerville}.
Knowing
the EBL as a function of redshift, one can calculate the optical depth with
respect to pair creation and determine the propagation length of
gamma-rays from cosmologically distributed sources such as Blazars
or Gamma Ray Bursts
\cite{lit:salsteck,lit:somerville}.

\section*{EBL-MODEL}

\begin{figure*}[t!] 
\centerline{\epsfig{file=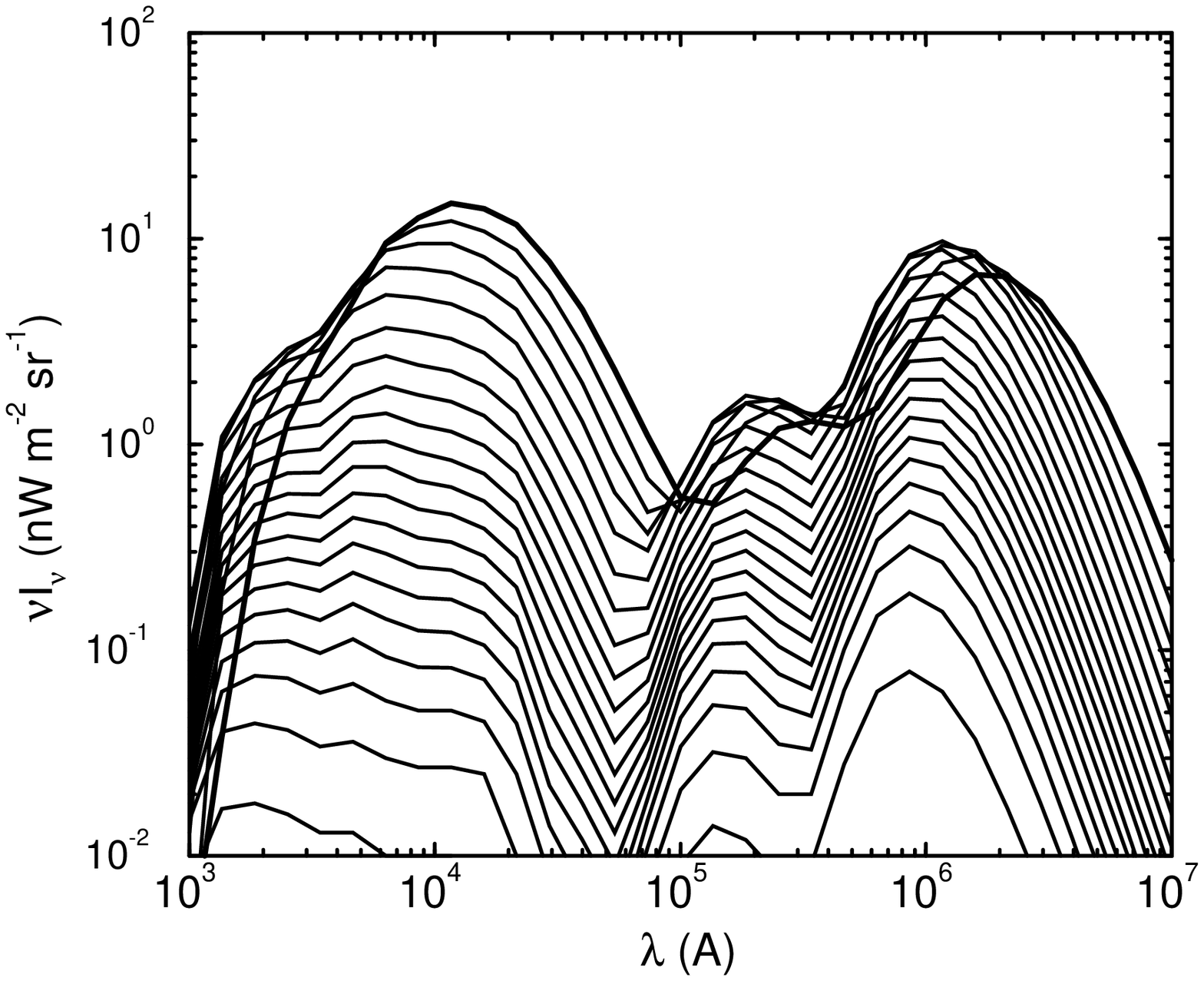,height=2.5in,width=3.0in}
\epsfig{file=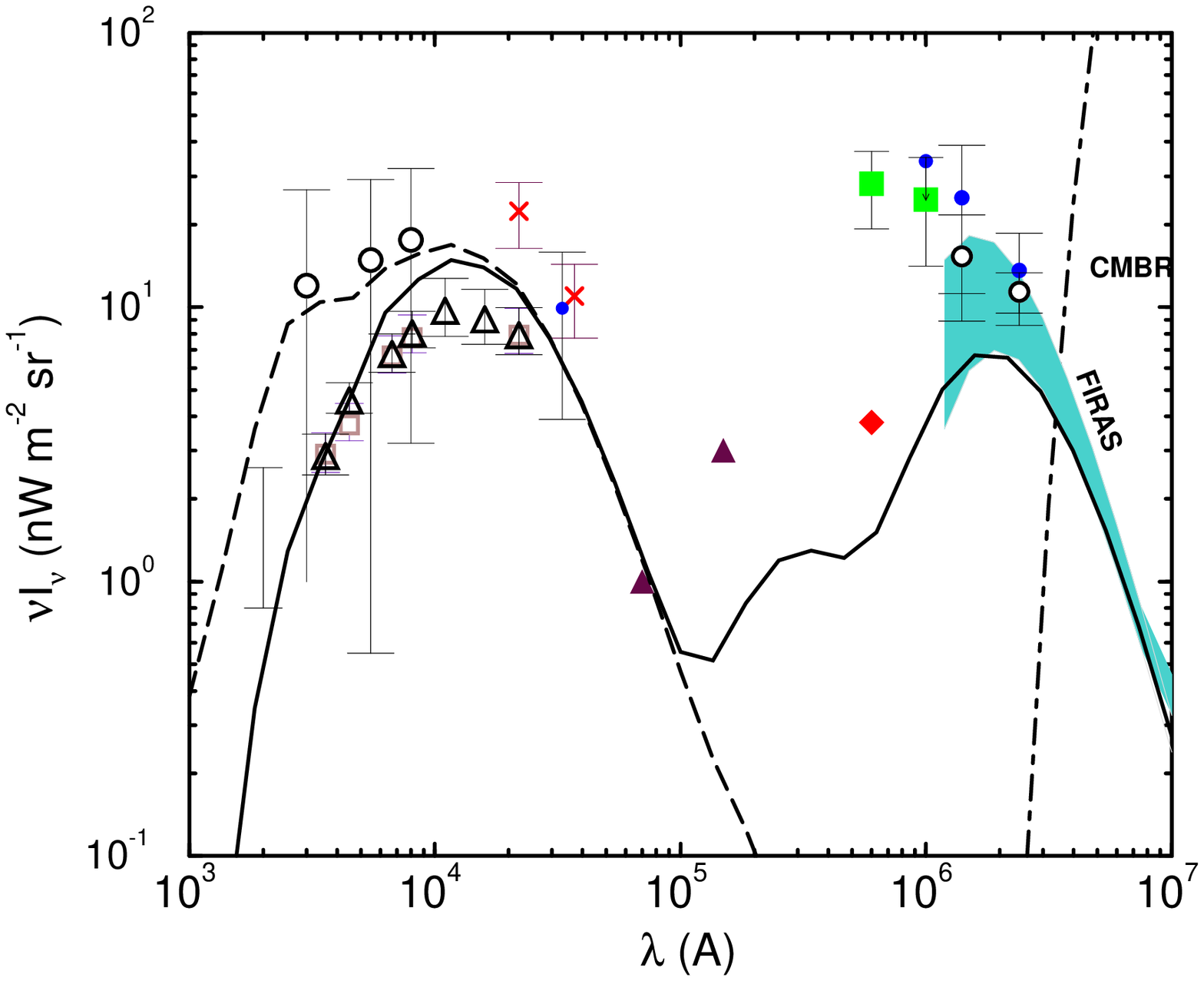,height=2.5in,width=3.0in}}
\vspace{5pt}
\caption{\emph{Left panel}: Evolution of EBL from $0<z<5$ $(\Delta z=0.2)$. 
($E_{B_V}$=0.14, $T_1$=150~K, $T_2$=35~K, $\alpha$=3.5, $\beta$=-1, $z_p$=1.3, $\dot{\rho_\ast}(z_p)$=0.15). 
\emph{Right panel}: Present-day EBL
with (\emph{solid line}) and without (\emph{long dashed line}) absorption. 
Data are Refs. from
[10](\emph{open triangles and squares}/HST), 
[11](\emph{circles}), 
[12](\emph{filled circle} at $3.5\cdot10^4$\AA /DIRBE), 
[13](''x''), 
[14](\emph{triangles}), 
[1](squares), 
[15](\emph{diamond}/IRAS, \emph{filled circles}/DIRBE), 
[16](\emph{open circles}), 
[17](\emph{shaded band}, FIRAS detection $1.25\cdot10^4$\AA-$50\cdot10^4$\AA).}
\label{fig2}
\end{figure*}

We start with the spectrum of an evolving simple stellar population (SSP)
\cite{lit:bruzual} 
(solar metalicity, Salpeter IMF
0.1-100 M$_\odot$). Next, we include absorption and re-emission
effects (dust and gas of ISM) using a SMC-like extinction law 
($E_{B-V}$)
and two modified blackbody spectra for the
re-emission
from dust ($T_1> T_2$). The Star Formation Rate (SFR) 
is $\dot{\rho_\ast}(z) \propto
(1+z)^{\alpha,\beta}$
with slopes $\alpha$ for $z\le z_p$ 
and $\beta$ for $z>z_p$, and a peak value 
$\dot{\rho_\ast}(z_p$).
By choosing proper
parameters, the emissivities $\epsilon_{\nu}(z)$ obtained from integrating the
SFR-weighted SSPs are brought to agreement with the data. A second
integration from 
$z$ to $z_{\rm max}$ yields the 
EBL comoving power spectrum 
(Fig.~\ref{fig2}) which is
independent of cosmology due to a cancellation of the cosmology-dependent
terms.
The EBL increases from high redshifts towards the present, but saturates at a 
roughly constant level beyond the peak in the  SFR. The EBL spectrum 
at high $z$ is dominated by young, UV-emitting massive stars.
70\% of the energy released by stars is re-emitted  by
dust in the infrared and far-infrared bands. The spectrum changes
towards its present shape as a result of a growing
admixture of
aged stars. About 40$\%$ of the integrated
present-day EBL is contained in the IR
(between $10^5${\AA}
and $10^7$\AA)(see Fig.\ref{fig2} and for details Ref.~\cite{lit:kneiske}).  

\section*{Gamma-Ray Attenuation}

\begin{figure*}[t!] 
\centerline{\epsfig{file=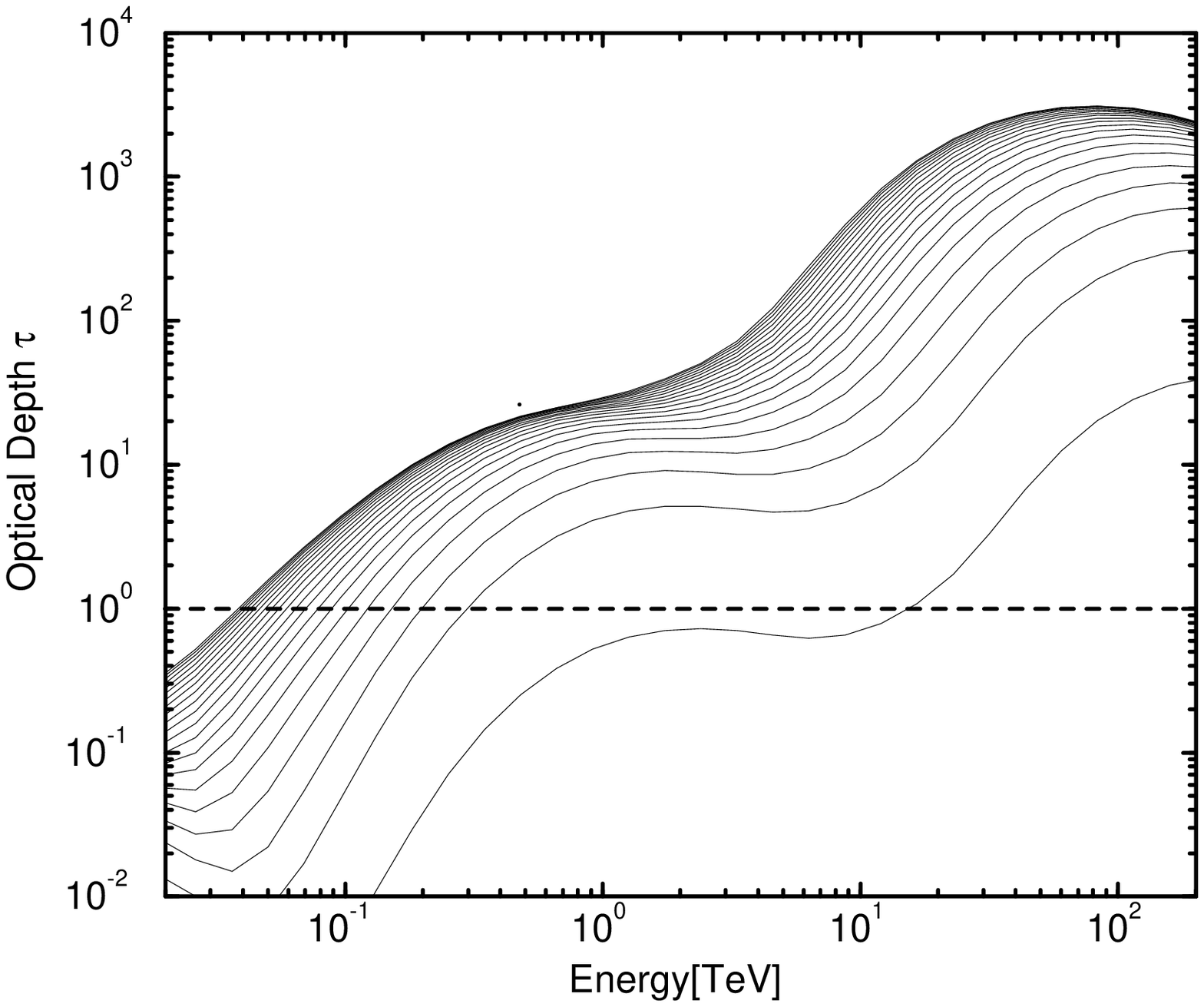,height=2.5in,width=3.0in}
\epsfig{file=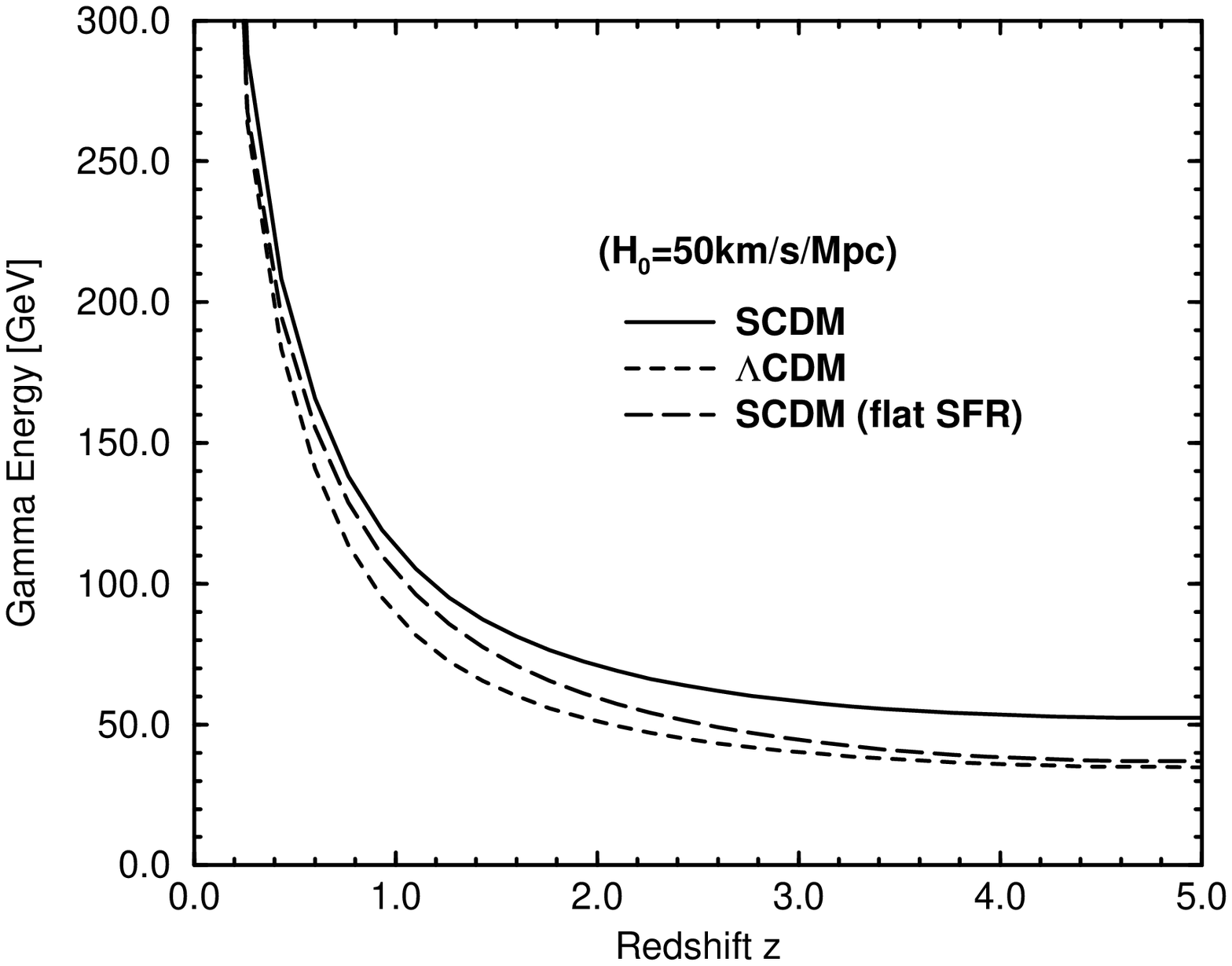,height=2.5in,width=3.0in}}
\vspace{10pt}
\caption{\emph{left panel}: Pair-creation optical depth for gamma-rays for
a range of redshifts (0.03$<$z$<$5). \emph{right panel}: Gamma ray horizon
defined by $\tau(z,E)=1$ for various models.}  \label{fig3}
\end{figure*}

The pair creation optical depth $\tau_{\gamma\gamma}(E,z)$
for gamma-rays at cosmological
redshift $z$ and 
observed energy $E$ is defined as
the ratio of
coordinate distance and mean free path
\cite{lit:salsteck}.
The optical depth evaluated for our EBL and a range of
redshifts ($0.03<z<4$) is shown in Fig.\ref{fig3} ($H_0=
50$~km~s$^{-1}$~Mpc$^{-1}$, $\Omega=1$, and $\Omega_\Lambda=0$).

\begin{figure*}[t!] 
\centerline{\epsfig{file=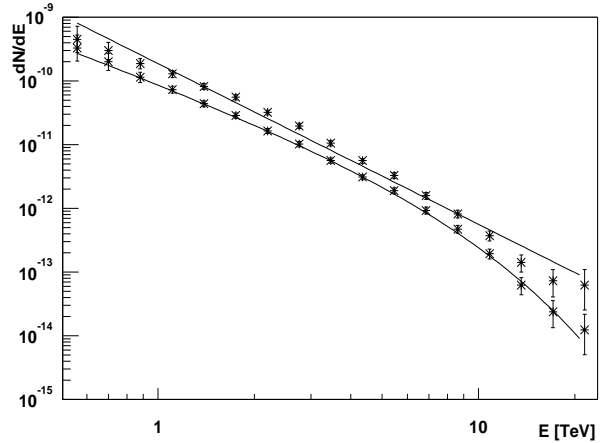,height=2.5in,width=3.5in}}
\vspace{10pt}
\caption{Intrinsic spectrum of Mkn501 inferred from our EBL model
({\it upper stars}) and fitted
by a power law with index
$2.49\pm0.04$ ({\it upper line}).  HEGRA original data 
from Ref.~[18]({\it lower stars})
and power law times exponential fit ({\it lower line}) 
from Ref.~[19] are shown
for comparison.}
\label{fig5} \end{figure*}

We can use the optical depth of gamma-rays to infer the intrinsic 
spectrum of Mkn501 from recent data.  This is achieved by
simply multiplying the data
with the factor $\exp[\tau_{\gamma \gamma}(0.034,E)]$, and the result is shown
in Fig.\ref{fig5}.

The gamma-ray horizon 
$\tau_{\gamma \gamma}(z,E)=1$ defines an interesting
relation between distance and cut-off
energy. In contrast to the EBL, the gamma-ray horizon depends on
cosmology through the integration over a cosmic line
element $dl/dz$. If we switch 
from a standard Cold Dark Matter model
(Fig.\ref{fig3}, {\it
right panel}, solid line) to one with a 
lower matter density (dashed line, $\Omega_\Lambda=0.7$),
the horizon moves closer and the number of sources which
can be detected  above a given gamma-ray threshold energy decreases. 
The effect could also be mimicked
by a flatter SFR
\cite{lit:steidel}.

\section*{Conclusions}
We have developed a simple model for the evolution of the extragalactic 
background radiation field using a minimum number of parameters by
fitting the emissivity predicted by the model to recent results
of high-redshift galaxy surveys.
Future improvements will include metalicity effects
and a PAH feature around 10$\mu$m. The intrinsic spectrum of
Mkn501 inferred from our model is in good agreement with theoretical
predictions (see Ref.~\cite{lit:mannheim} for a discussion) rendering
claims for an extremely high level of the infrared background light 
intensity implausible. The EBL model is a powerful
tool to analyze gamma-ray spectra from high-redshift sources
in the 10~GeV to
TeV regime for which the pair attenuation occurs in
interactions with ultraviolet-to-infrared photons allowing to
probe the era of maximum star formation or cosmology.

{\bf Acknowledgements:} We acknowledge support under DESY-HS/AM9MGA7 and
thank  H.~Krawcinsky, A.~Konopelko, 
J.~Primack, F.W.~Stecker, and D.~Lemke for discussions.


\begin{references}
\bibitem{lit:finkbeiner} Finkbeiner,D.P., Davis,M., Schlegel,D.J., astro-ph/0004175(2000).
\bibitem{lit:salsteck}Salamon, M.H., \& Stecker, F.W., {\it ApJ} {\bf 493}, 547(1998).
\bibitem{lit:malsteck} Malkan, M. \& Stecker, F.W., {\it ApJ} {\bf 496}, 13(1998).
\bibitem{lit:kneiske1}Kneiske,T.M., Mannheim,K., astro-ph/9912450(1999). 
\bibitem{lit:somerville}Somerville, R. \& Primack, J.R., {\it MNRAS} {\bf 310}, 1087(1999). 
\bibitem{lit:bruzual}Bruzual,A.G., \& Charlot,S., {\it ApJ} 
{\bf 405}, 538(1993); Charlot, {\it priv. comm.}.
(1999).
\bibitem{lit:kneiske}Kneiske, T.M., Mannheim, K. \& Hartmann, D., submitted (2000). 
\bibitem{lit:steidel} Steidel, C.C. et al.,  {\it ApJ} {\bf 519}, 1 (1999).
\bibitem{lit:mannheim}Mannheim, K., {\it Science} {\bf 279}, 684 (1998).
\bibitem{lit:madau98} Pozzetti, L. et al., {\it MNRAS} {\bf 298}, 1133(1998);
Madau,P. \& Pozzetti,L., {\it MNRAS} {\bf 312}, L9(2000).
\bibitem{lit:bernstein} Bernstein, R.A., Freeman, W.L. \& Madore, B.F., submitted (2000).
\bibitem{lit:dwek}Dwek, E., Arendt, R.G. {\it ApJ} {\bf 508}, 9(1998). 
\bibitem{lit:gorjan}Gorjian, V., Wright, E.L. \& Chary, R.R., {\it ApJ} {\bf 536} 550(2000). 
\bibitem{lit:hacking}Altieri, B. {\it A\&A} {\bf 343}, 65(1998).
\bibitem{lit:hauser}Hauser, M.G. et al., {\it ApJ} {\bf 508}, 25(1998).
\bibitem{lit:lagache}Lagache et al., {\it ApJ} {\bf 344}, 322(1999).
\bibitem{lit:fixsen}Fixsen et al., {\it ApJ} {\bf 508}, 123(1998).
\bibitem{lit:aha}Aharonian et al., {\it A\&A} {\bf 349}, 11(1999).
\bibitem{lit:kono}Konopelko et al., {\it ApJ} {\bf 518}, 13(1999).


\end{references}
\end{document}